\documentclass[a4paper,11pt]{article}
\usepackage{pos}
\usepackage{hyperref}

\title{Recent open heavy flavor studies for the Electron-Ion Collider}

\author*[a]{Xuan Li}

\affiliation[a]{Los Alamos National Laboratory,\\
  Los Alamos, USA}

\emailAdd{xuanli@lanl.gov}

\abstract{The future Electron-Ion Collider (EIC) will operate a series of high-luminosity high-energy electron+proton ($e+p$) and electron+nucleus ($\textit{e + A}$) collisions to study several fundamental questions in the high energy and nuclear physics field. Heavy flavor hadron and jet production at the EIC plays an important role in exploring both potential modification on the initial-state nuclear parton distribution functions (nPDFs) and final-state parton propagation and hadronization processes under different nuclear medium conditions. The current design of the EIC ePIC detector has good performance of vertex and track reconstruction, particle identification and energy determination in the pseudorapidity region of $-3.5<\eta<3.5$, which will enable a series of high precision heavy flavor hadron and jet measurements. Latest simulation studies of the projected nuclear modification factor $R_{eA}$ of heavy flavor jets and heavy flavor hadron inside jets in $e+p$ and $\textit{e + Au}$ collisions at $\sqrt{s} =$ 28.6 GeV and 63.2~GeV as well as the projected statistical accuracy of inclusive and differential charm baryon over meson ratio measurements in $e+p$ collisions will be presented. The impacts of these proposed EIC measurements on constraining the heavy quark propagation properties in cold nuclear medium and exploring the heavy quark hadronization process will be discussed.}

\FullConference{10th International Conference on Quarks and Nuclear Physics (QNP2024)\\
8-12 July, 2024\\
Barcelona, Spain\\}

\begin{document}
\maketitle

\section{Introduction}
\label{sec1}
The future Electron-Ion Collider (EIC), which is scheduled to start construction in 2025, will run $e+p$ and $\textit{e + A}$ collisions at the center of mass energies of $20-141$~GeV \cite{eic_YR}. Both the electron and proton beam can be operated in polarized and unpolarized modes. The species of the nuclear beam vary from deuteron to uranium with the mass number changing from 2 to 238. The instantaneous luminosity of the EIC can reach $10^{33-34}~\rm{cm}^{-2}\rm{s}^{-1}$, which is nearly 3 orders of magnitude higher than the last $e+p$ collider: HERA. With these high-luminosity $e+p$ and $\textit{e + A}$ collisions, the future EIC will be able to perform a series of high precision measurements to study the 3D structure of nucleon/nuclei, help address the proton spin puzzle, probe the nucleon/nuclei parton density extreme (e.g., gluon saturation) and explore how quarks and gluons form visible matter under different medium conditions. The EIC will support up to two interaction points: IP6 and IP8. The detector locates at IP6, is under design by the ePIC collaboration and is also referred to as the ePIC detector. To enhance the delivery luminosity at the interaction region and mitigate beam backgrounds, a 25~mrad crossing angle has been introduced to IP6.

The ePIC detector consists of optimized vertex, tracking, particle identification (PID), electromagnetic calorimeter (EMCal) and hadronic calorimeter (HCal) subsystems, which enables high precision hadron and jet measurements in the pseudorapidity region of $-3.5<\eta<3.5$. The central ePIC detector spans an area of 9.5~m longitudinally by 3.3~m horizontally. The ePIC detector also includes the far-forward and far-backward subsystems to monitor the beam luminosity, detect nuclear breakup and measure the exclusive processes. Extensive simulation studies have been carried out to evaluate the performance of the ePIC detector. The tracking momentum and spatial resolutions, the PID separation capabilities and the electromagnetic and hadronic energy resolutions of the current ePIC detector design meet the detector requirement in the EIC yellow report \cite{eic_YR} in most kinematic regions. Recent open heavy flavor hadron and jet simulation studies with the current ePIC detector performance will be presented, and their physics impacts in exploring the heavy quark parton energy loss mechanism and hadronization process will be discussed.

\section{Open heavy flavor hadron and jet reconstruction in $e+p$ simulation}
\label{sec2}
\begin{figure}[ht]
\centering
\includegraphics[width=0.32\textwidth,clip]{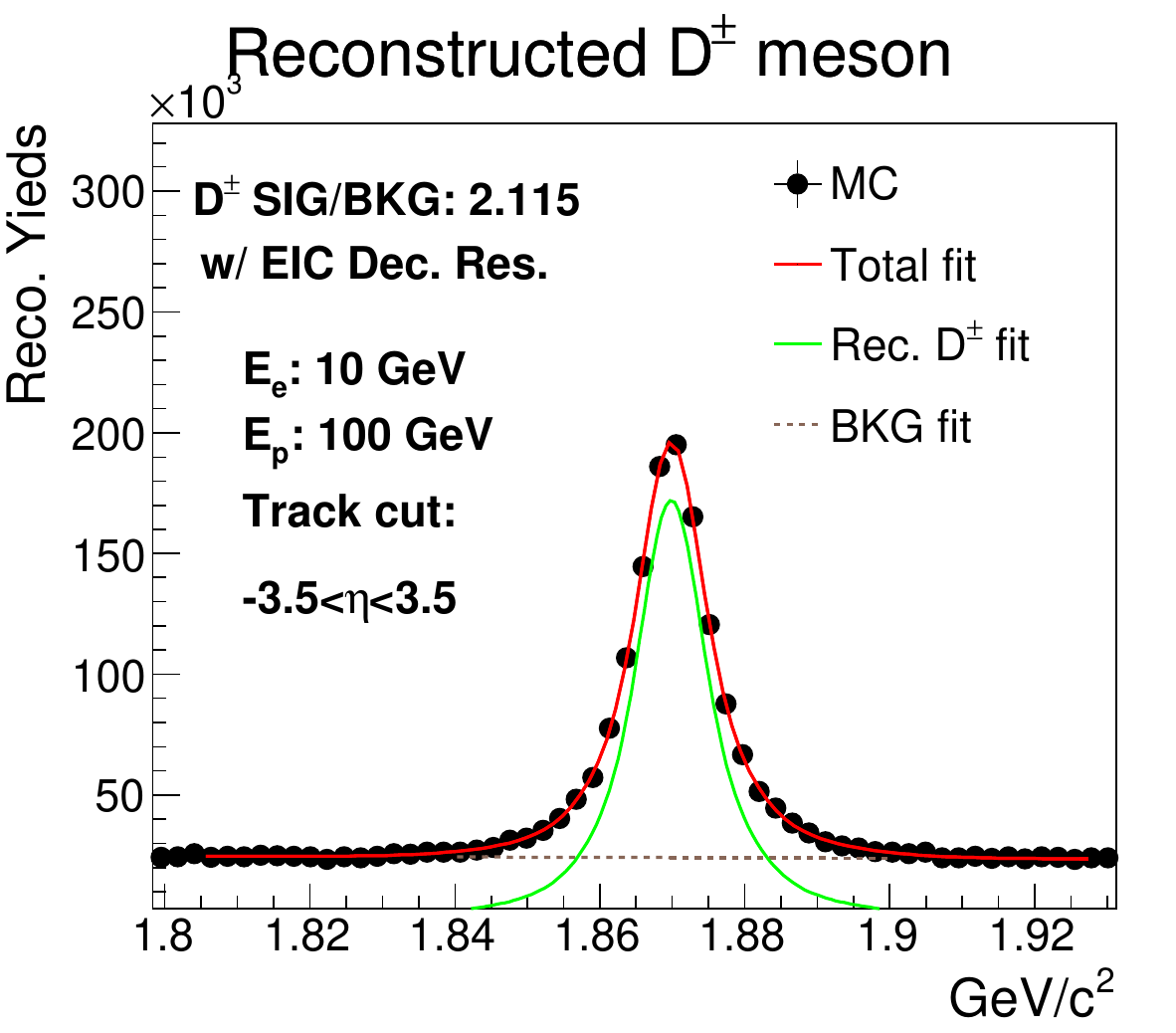}
\includegraphics[width=0.32\textwidth,clip]{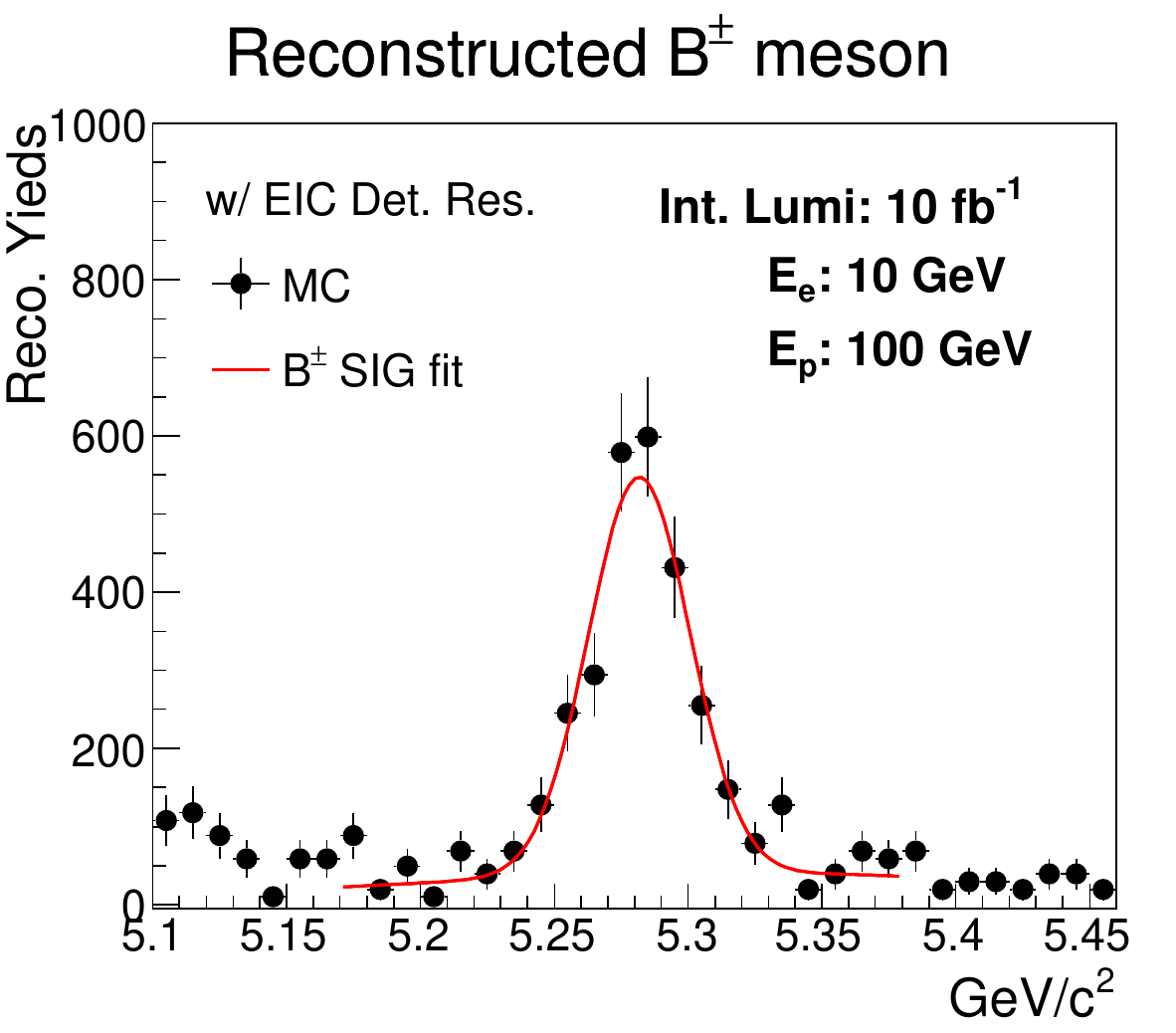}
\includegraphics[width=0.32\textwidth,clip]{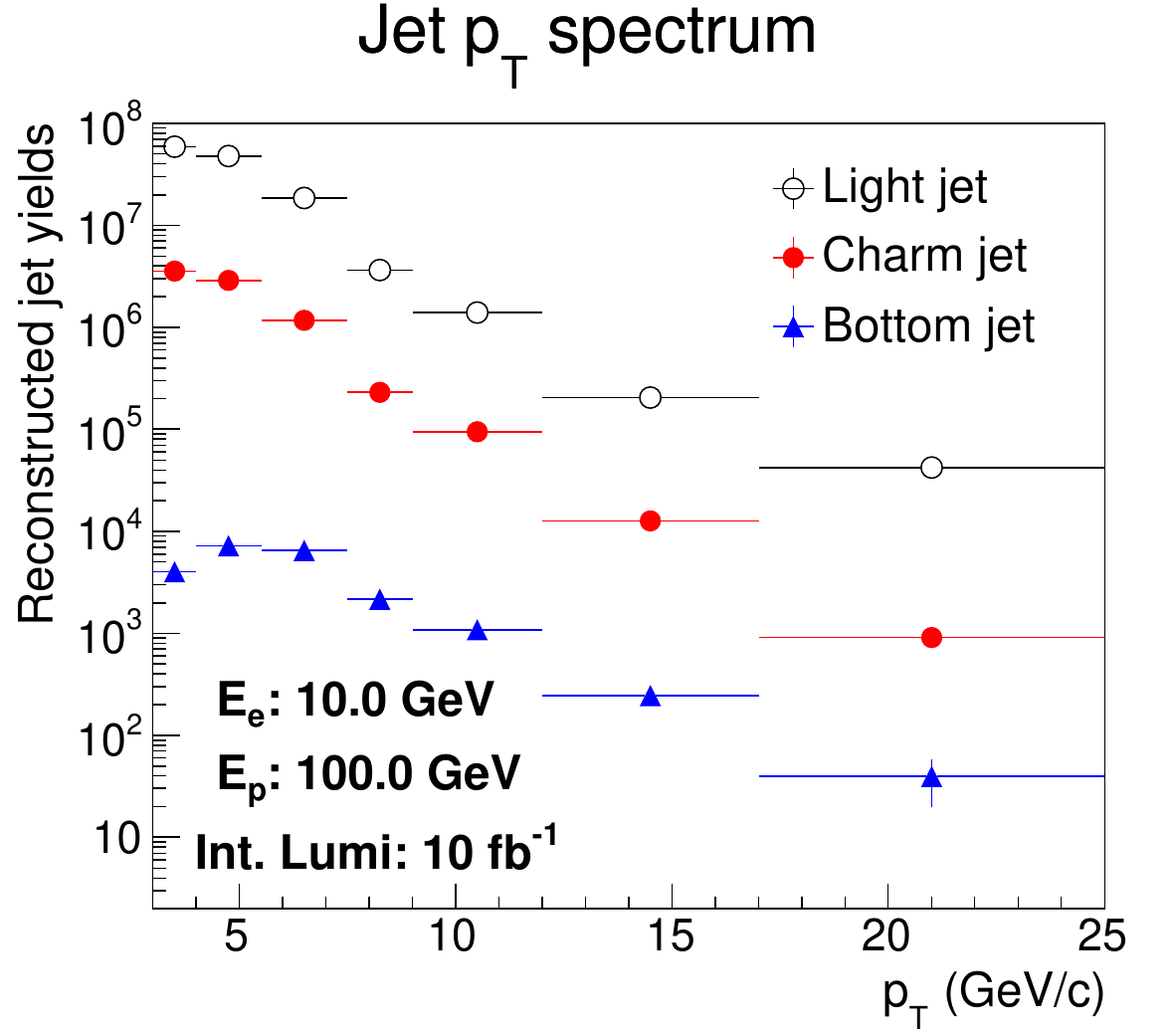}
\caption{Left: reconstructed $D^{\pm}$ mass spectrum in 63.2~GeV $e+p$ simulation. Middle: the invariant mass spectrum of reconstructed $B^{\pm}$ in 63.2~GeV $e+p$ simulation. Right: the $p_{T}$ spectrum of reconstructed light-jets (black open circles), charm-jets (red closed circles) and bottom-jets (blue closed triangles) in 63.2~GeV $e+p$ simulation. The parameterized ePIC detector performance has been utilized in these simulation studies.}
\label{fig:reco}       
\end{figure}

A standalone simulation framework, which is based on the Deeply Inelastic Scattering (DIS) $e+p$ event generation in PYTHIA8 \cite{py8}, has been established for heavy flavor hadron and jet studies at the EIC. The parameterized ePIC detector response derived from recent GEANT4 \cite{geant4} simulations has been included in this framework by smearing the true particle information. A series of heavy flavor hadron reconstruction studies have been performed in the simulation to validate the ePIC detector capability. The left panel of Figure~\ref{fig:reco} shows the invariant mass spectrum of reconstructed $D^{\pm}$ through the decay channel of $D^{\pm} \rightarrow K^{\mp}\pi^{\pm}\pi^{\pm}$ in a simulation for 10 GeV electron and 100 GeV proton collisions. The mass spectrum of reconstructed $B^{\pm}$ through the decay channel of $B^{\pm} \rightarrow J/\psi (\rightarrow l^{+}l^{-}) + K^{\pm}$ in 63.2~GeV $e+p$ simulation is shown in the middle panel of Figure~\ref{fig:reco}. The good performance of ePIC vertex, tracking, PID and calorimeter subsystems will enable precise jet measurements as well. The right panel of Figure~\ref{fig:reco} illustrates the reconstructed jet $p_{T}$ distributions for light-, charm- and bottom-jets in 63.2~GeV $e+p$ simulation. Due to the low particle multiplicity in $e+p$ and $\textit{e + A}$ collisions at the EIC energies, jets are reconstructed with the anti-$k_{T}$ algorithm and the cone radius is selected at $R=1.0$. The flavor of jets is tagged according to the characteristics of reconstructed decay/secondary vertex inside jets. The majority of reconstructed jets in 63.2~GeV $\textit{e + p/A}$ collisions has the low $p_{T}$ coverage of $5-20$~GeV/c and the jet $p_{T}$ coverage is lower in 28.6~GeV $e+p/A$ collisions.

\section{Heavy flavor jet $R_{eAu}$ projection at the EIC}
\label{sec3}
\begin{figure}[ht]
\centering
\includegraphics[width=0.47\textwidth,clip]{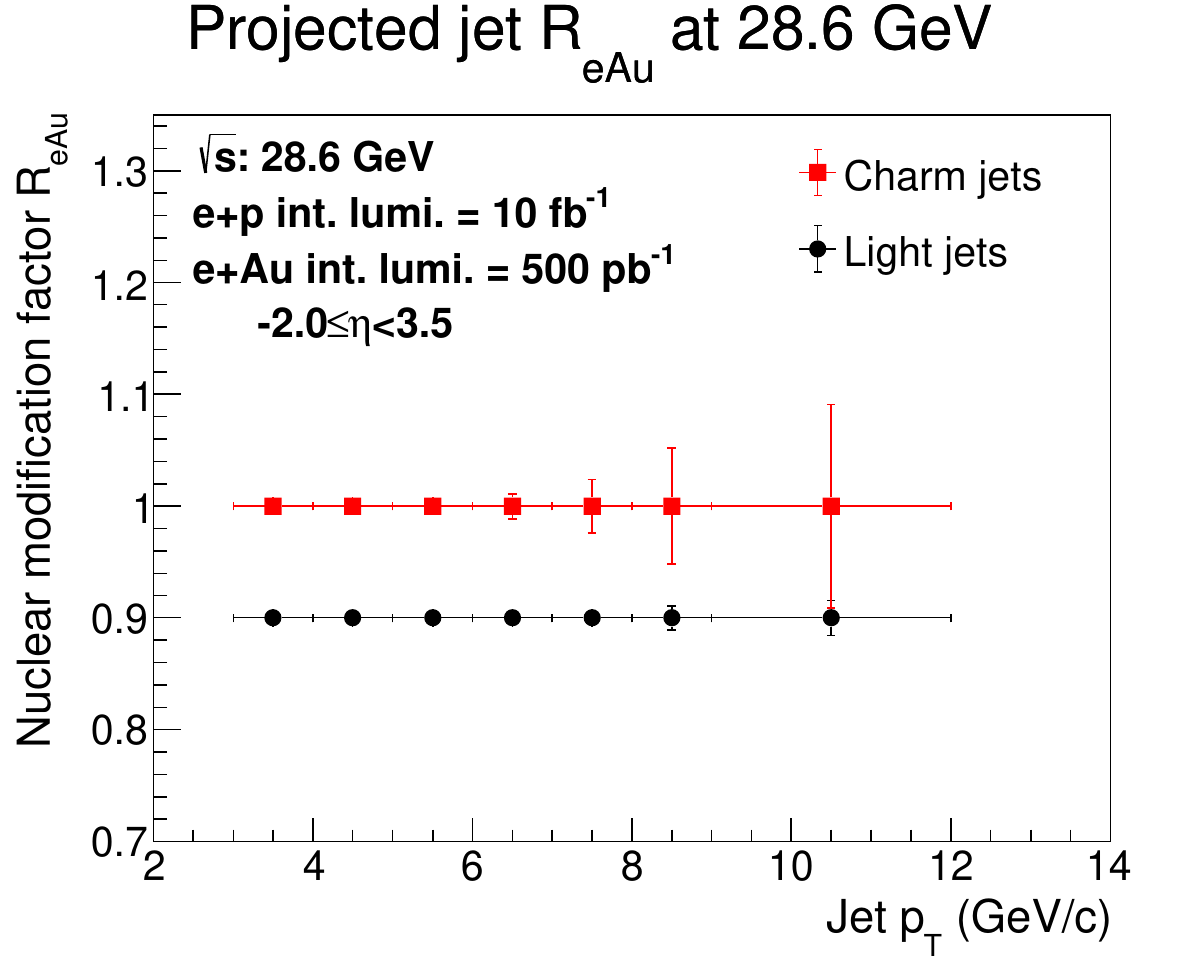}
\includegraphics[width=0.47\textwidth,clip]{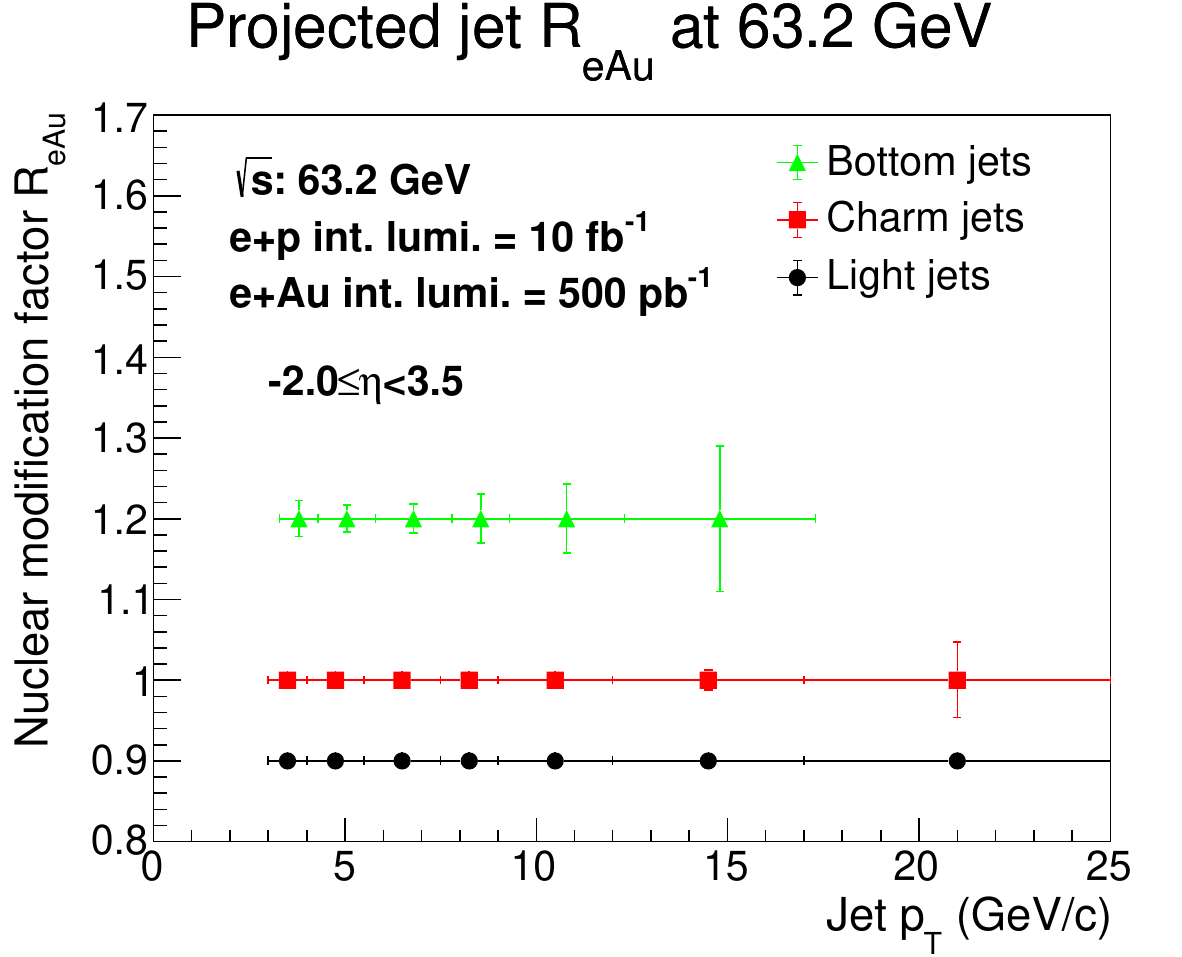}
\caption{The projected statistical accuracy of $p_{T}$ dependent nuclear modification factor $R_{eAu}$ for reconstructed jets with different flavors in $\textit{e + Au}$ collisions at 28.6~GeV (left) and 63.2~GeV (right). Jets are reconstructed within the pseudorapidity region of $-2.0<\eta<3.5$ using the anti-$k_{T}$ algorithm with the cone radius R = 1.0. The $e+p$ integrated luminosity is 10~$fb^{-1}$ and the $\textit{e + Au}$ integrated luminosity is 500~$pb^{-1}$.}
\label{fig:jet_ReA}       
\end{figure}

The low $p_{T}$ heavy flavor jets to be measured at the EIC are an ideal probe to systematically explore the flavor dependent energy loss mechanism under different medium conditions from heavy ion measurements at Relativistic Heavy Ion Collider (RHIC) and the Large Hadron Collider (LHC). Moreover, the kinematics of the hard scattering partons can be precisely determined in the DIS process. One typical measurement to study the heavy quark energy loss process is the $p_{T}$ dependent heavy flavor jet nuclear modification factor $R_{eA}$, which is defined as the ratio of the cross section measured in $\textit{e + A}$ collisions over that in $e+p$ collisions and scaled by the mass number A. Figure~\ref{fig:jet_ReA} presents the projected statistical accuracy of $p_{T}$ dependent nuclear modification factor $R_{eAu}$ for reconstructed light and heavy flavor jets in $\textit{e + Au}$ collisions at 28.6~GeV (left) and 63.2~GeV (right). Great precision can be achieved by future EIC light- and charm-jet $R_{eAu}$ measurements with $p_{T}<10$~GeV/c in 28.6~GeV $e+p$ and $\textit{e + Au}$ collisions. The $p_{T}$ coverage of light-, charm- and bottom-jet $R_{eAu}$ measurements can be extended to $p_{T}<17$~GeV/c in 63.2~GeV $e+p$ and $\textit{e + Au}$ collisions. All these studies will significantly improve our understanding about the interplay of collisional and radiative energy loss for heavy quarks at the early phase of collisions.

\section{Charm baryon over meson ratio studies at the EIC}
\label{sec4}
\begin{figure}[ht]
\centering
\includegraphics[width=0.47\textwidth,clip]{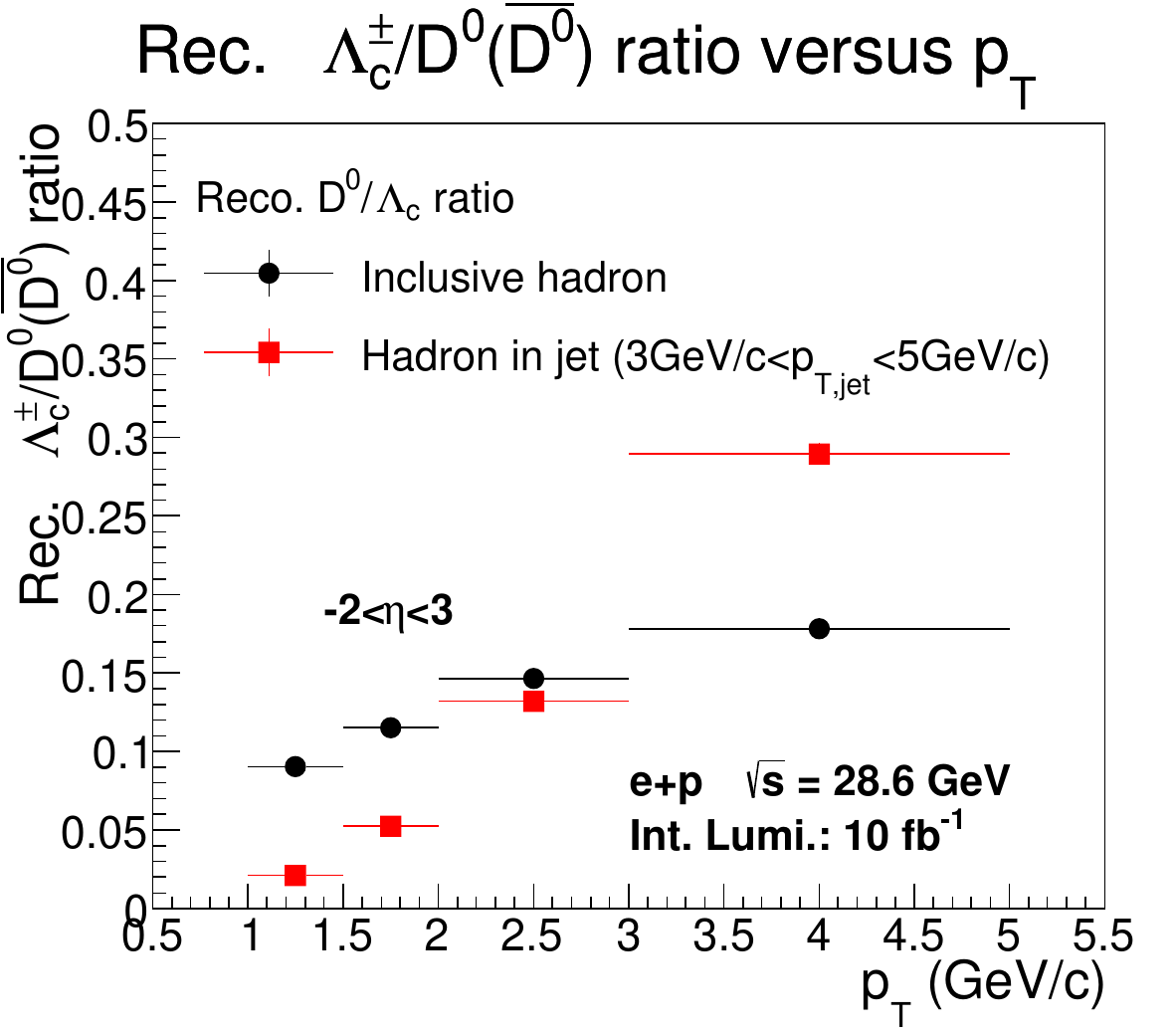}
\includegraphics[width=0.47\textwidth,clip]{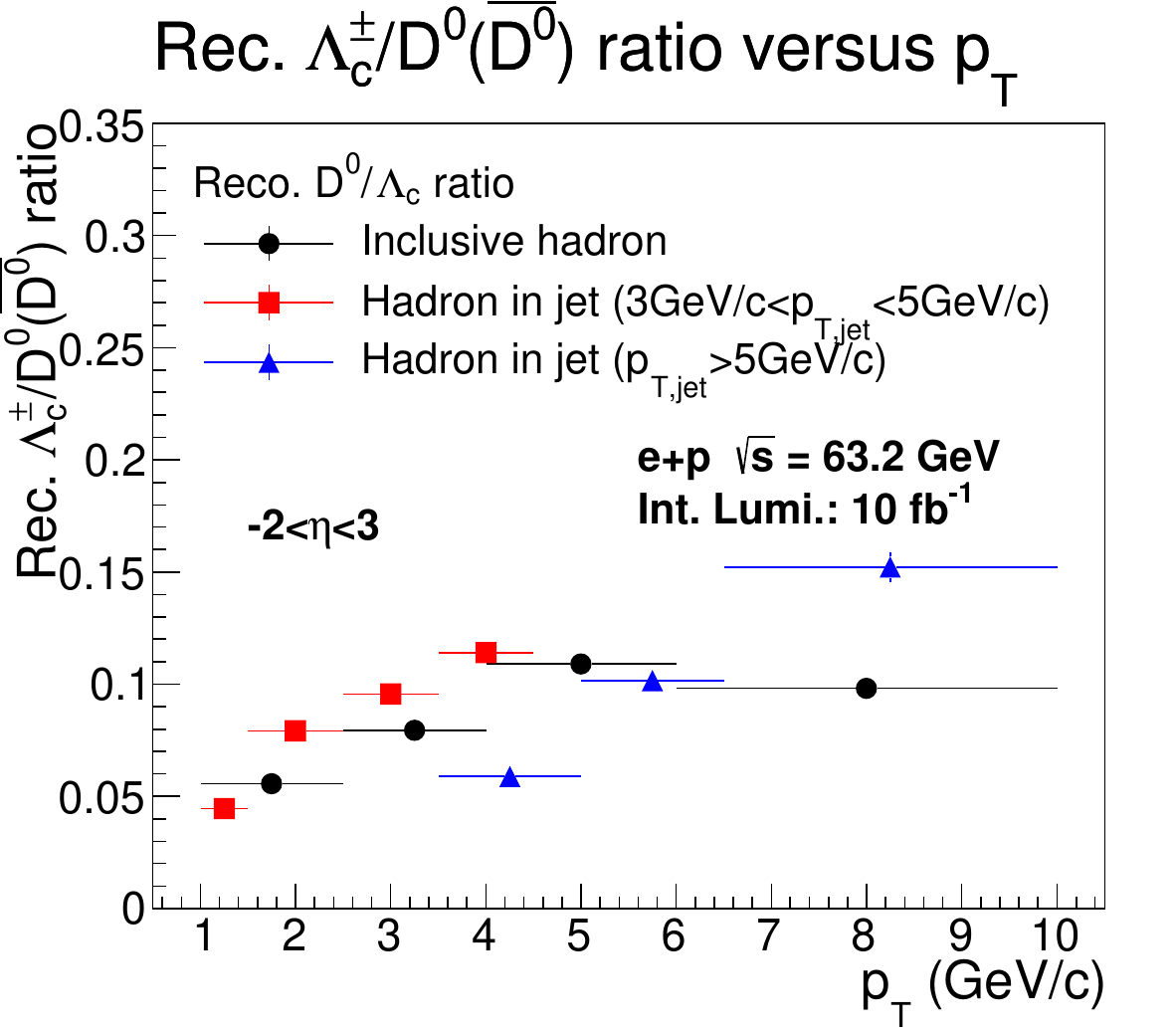}
\caption{The hadron $p_{T}$ dependent reconstructed $\Lambda_{c}^{\pm}/D^{0}$($\bar{D^{0}}$) ratio projections in $e+p$ simulation at 28.6~GeV (left) and 63.2~GeV (right). Comparison of the charm baryon over meson ratio which is measured inclusively with that when reconstructed $\Lambda_{c}^{\pm}$ and $D^{0}$($\bar{D^{0}}$) are associated with low $p_{T}$ jets (3~GeV/c < $p_{T,jet}$ < 5~GeV/c) has been performed in both collision systems. In addition, the ratio of reconstructed $\Lambda_{c}^{\pm}$ inside jet over reconstructed $D^{0}$($\bar{D^{0}}$) inside jet with $p_{T,jet}$ > 5~GeV/c has been studied in 63.2~GeV $e+p$ simulation.}
\label{fig:chad_ratio}       
\end{figure}

The universality of charm quark fragmentation function was challenged by recent inclusive charm baryon over meson ratio results in $p+p$, p+A and A+A collisions at RHIC and the LHC (e.g., $\Lambda_{c}/D^{0}$). Precise reconstruction of $D^{0}$ ($\bar{D^{0}}$) and $\Lambda_{c}^{\pm}$ within and without a associated jet has been validated in $e+p$ simulation for the EIC \cite{cr_jet_eic}. Figure~\ref{fig:chad_ratio} compares the hadron $p_{T}$ dependent inclusive $\Lambda_{c}^{\pm}/D^{0}$($\bar{D^{0}}$) projection in simulation with the same observable when $\Lambda_{c}^{\pm}$ and $D^{0}$($\bar{D^{0}}$) hadrons are associated with jets for 28.6~GeV (left) and 63.2~GeV (right) $e+p$ collisions respectively. Different phase spaces of the charm quark fragmentation process can be accessed by measuring charm hadrons inside jets in different $p_{T,jet}$ regions. Direct information of charm quark differential fragmentation functions will be able to be extracted from future EIC charm baryon over meson ratio measurements with different selections. Further simulation studies will be expanded to similar measurements in different $\textit{e + A}$ collisions to evaluate potential differences between charm quark to meson fragmentation functions and charm quark to baryon fragmentation functions under different cold nuclear medium conditions. 

\section{Heavy flavor hadron inside jet $R_{eAu}$ projection at the EIC}
\label{sec5}
\begin{figure}[ht]
\centering
\includegraphics[width=0.7\textwidth,clip]{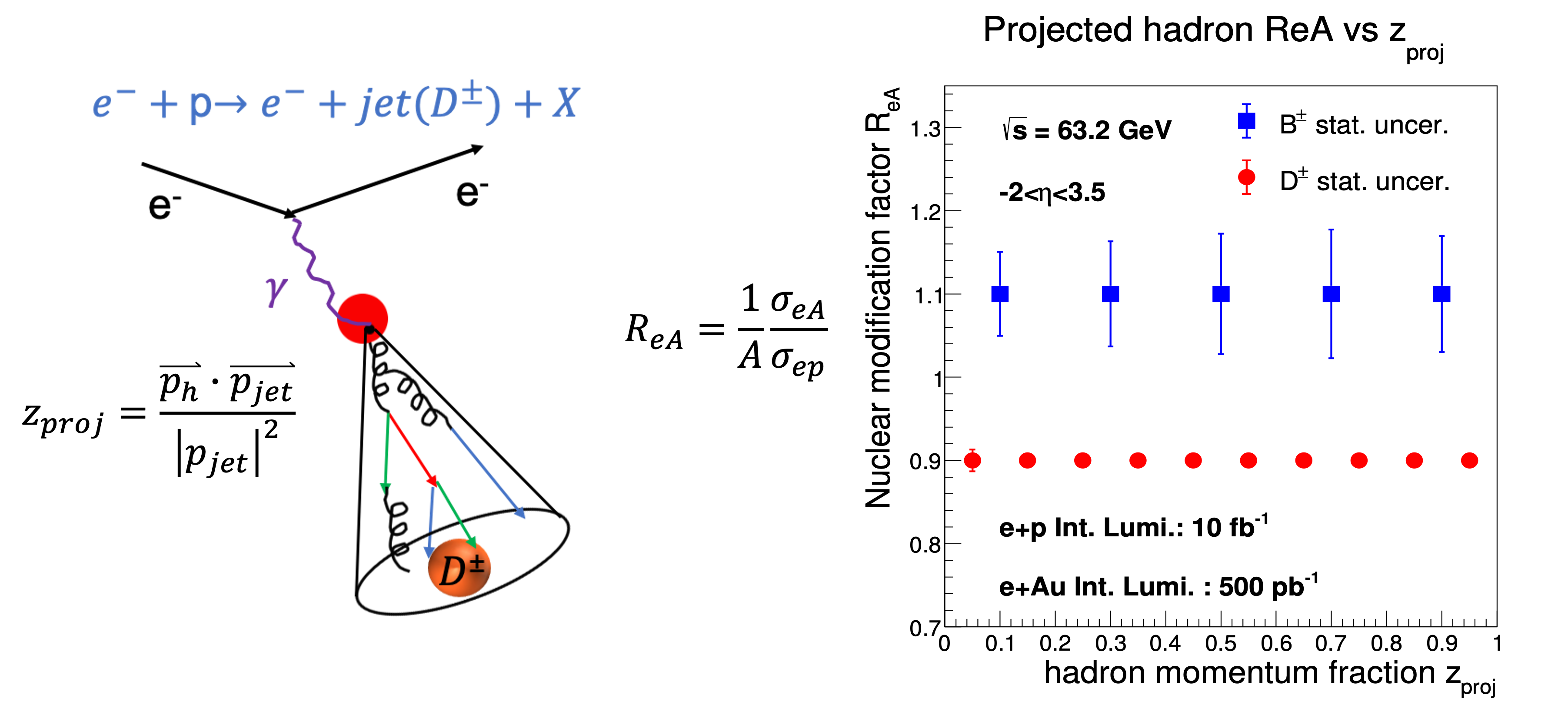}
\caption{Left: The schematics of the $D^{\pm}$ meson production inside jet in $e+p$ collisions. Right: The projected statistical accuracy of hadron momentum fraction $z_{proj}$ ($z_{proj} = {\vec{p_{h}}} \cdot {\vec{p_{jet}}}/|p_{jet}|^{2}$) dependent $R_{eAu}$ for reconstructed $D^{\pm}$ inside jets (closed red circles) and reconstructed $B^{\pm}$ inside jets (closed blue rectangulars) in 63.2~GeV $e+p$ and $\textit{e + Au}$ collisions. The $e+p$ integrated luminosity is 10~$fb^{-1}$ and the $\textit{e + Au}$ integrated luminosity is 500~$pb^{-1}$.}
\label{fig:hfh_ReA}       
\end{figure}

\begin{figure}[ht]
\centering
\includegraphics[width=0.96\textwidth,clip]{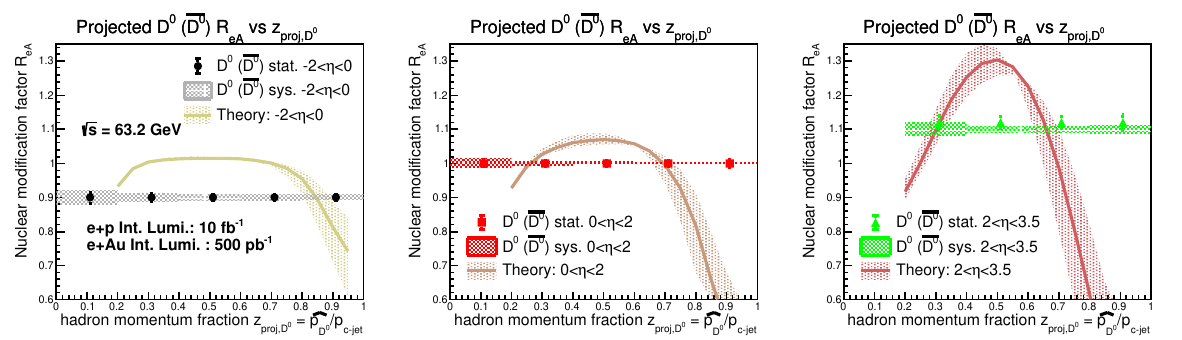}
\caption{The projected statistical and systemic uncertainties of the $z_{proj}$ dependent $R_{eA}$ of reconstructed $D^{0}$ ($\bar{D^{0}}$) inside charm-jet with the evaluated ePIC detector performance in comparison with theoretical predications based on the parton energy loss model \cite{hf_th} within three pseudorapidity bins in 63.2~GeV $\textit{e + Au}$ collisions. The pseudorapidity coverage of reconstructed $D^{0}$ ($\bar{D^{0}}$) and charm-jets from the left panel to the right panel is $-2<\eta<0$, $0<\eta<2$ and $2<\eta<3.5$. The $e+p$ integrated luminosity is 10~$fb^{-1}$ and the $\textit{e + Au}$ integrated luminosity is 500~$pb^{-1}$.}
\label{fig:Dj_ReA}       
\end{figure}

Heavy flavor hadron inside jet measurements at the EIC will achieve good sensitivity to explore the heavy quark hadronization process. The hadron momentum fraction, $z_{proj}$ ($z_{proj} = {\vec{p_{h}}} \cdot {\vec{p_{jet}}}/|p_{jet}|^{2}$), which is defined as the relative momentum fraction carried by a hadron relative to the associated jet axis, is a key kinematic variable in determining fragmentation functions. In addition to inclusive heavy hadron and jet measurements as discussed in Sec.~\ref{sec2} and \ref{sec3}, a series of inclusive and differential heavy flavor hadron inside jet studies will be performed in $e+p$ and $\textit{e + Au}$ collisions at the EIC. The projected statistical accuracy of the hadron momentum fraction $z_{proj}$ dependent nuclear modification factor $R_{eAu}$ of reconstructed $D^{\pm}$ inside jets and reconstructed $B^{\pm}$ inside jets with the parameterized ePIC detector performance in 63.2~GeV $e+p$ and $\textit{e + Au}$ collisions. Better than $8\%$ statistical uncertainties can be achieved by reconstructed $B^{\pm}$ inside jet $R_{eAu}$ measurements with 10~$fb^{-1}$ $e+p$ integrated luminosity and 500~$pb^{-1}$ $\textit{e + Au}$ integrated luminosity. The great statistics of reconstructed $D^{0}$ ($\bar{D^{0}}$) inside jet measurements at the EIC will allow further differential measurements to explore the charm quark hadronization process. Figure~\ref{fig:Dj_ReA} presents the projected statistical and systemic uncertainties of the $z_{proj}$ dependent $R_{eA}$ of reconstructed $D^{0}$ ($\bar{D^{0}}$) inside charm-jet with the ePIC detector performance in the pseudorapidity regions of $-2<\eta<0$ (left), $0<\eta<2$ (middle) and $2<\eta<3.5$ (right) in 63.2~GeV $\textit{e + Au}$ collisions. Comparison with the Next-to-Leading Order (NLO) parton energy loss theoretical predictions indicates better precision will be achieved by the future EIC $D^{0}$ ($\bar{D^{0}}$) inside jet $R_{eAu}$ measurements especially in the large $z_{proj}$ region at forward pseudorapidity, in which a relatively significant uncertainty exists in the current extracted charm quark fragmentation function. Moreover, great discriminating power will be offered by these EIC heavy flavor hadron inside jet $R_{eA}$ measurements to separate the scenario when the hadronization happens outside the nucleus according to the parton energy loss model from that when the hadronization happens inside the nucleus based on the absorption model assumption.

\section{Summary and Outlook}
\label{sec:sum}
The EIC is scheduled to start construction in 2025 and it is expected to be operated in early 2030s. A series of high-precision heavy flavor hadron/jet and heavy flavor hadron inside jet measurements will be carried out in high-luminosity $e+p$ and $\textit{e + A}$ collisions at different center-of-mass energies at the future EIC to explore both the initial-state and final-state nuclear effects. The wide pseudorapidity coverage of the ePIC detector will enable good heavy flavor di-jet and heavy flavor jet substructure measurements as well. These studies will not only compensate the kinematic coverage gaps from previous and current results in different collision systems, but also provide better constraints in exploring the flavor dependent parton energy loss and hadronization processes under different nuclear medium conditions. Furthermore, additional information will be provided by these EIC heavy flavor studies in improving the understanding of current and future heavy ion measurements at RHIC and the LHC.

\section{Acknowledgements}
This work is primarily supported by the Los Alamos National Laboratory LDRD project. The EIC physics standalone simulation studies use the parameterized outputs of recent detector performance evaluation carried out by the ePIC collaboration. We would like to express our appreciation to our theorist friends: Ivan Vitev, Weiyao Ke and Zhongbo Kang for various valuable feedback and suggestions.

\end{document}